\documentclass[11pt]{article}
\usepackage{graphicx}
\usepackage{amsmath}
\usepackage{natbib}

\begin{document}

\title{{\it Kepler}-16b: safe in a resonance cell}

\author{Elena A. Popova, Ivan I. Shevchenko \\ {\small Pulkovo Observatory of the Russian Academy of Sciences,} \\ {\small Pulkovskoje ave. 65, St.~Petersburg 196140, Russia}}

\maketitle

\begin{abstract}
The planet {\it Kepler}-16b is known to follow a circumbinary
orbit around a system of two main-sequence stars. We construct
stability diagrams in the ``pericentric distance~---
eccentricity'' plane, which show that {\it Kepler}-16b is in a
hazardous vicinity to the chaos domain~--- just between the
instability ``teeth'' in the space of orbital parameters. {\it
Kepler}-16b survives, because it is close to the stable
half-integer 11/2 orbital resonance with the central binary, safe
inside a resonance cell bounded by the unstable 5/1 and 6/1
resonances. The neighboring resonance cells are vacant, because
they are ``purged'' by {\it Kepler}-16b, due to overlap of
first-order resonances with the planet. The newly discovered
planets {\it Kepler}-34b and {\it Kepler}-35b are also safe inside
resonance cells at the chaos border.
\\
\\
\noindent {\it Keywords:} planets and satellites: general --- dynamical evolution
and stability: general --- formation: individual --- {\it
Kepler}-16, {\it Kepler}-16b
\end{abstract}

\section{Introduction}

A circumbinary planet in the {\it Kepler}-16 system was discovered
in 2011 basing on the data from the {\it Kepler} spacecraft
\citep{D11}. Its orbital parameters were determined in
\citep{D11,W12}. An application of an empirical
numerical-experimental criterion \citep{HW99} of the stability of
planetary circumbinary orbits, as well as long-term numerical
integrations of the {\it Kepler}-16b orbit, identifies this system
as stable, though not far (in about 10--20\% in the semimajor axis
of the planet) from the inner instability zone \citep{D11,W12}.
However, a fine structure of the stability border in the space of
orbital parameters has not yet been studied.

Although the planet {\it Kepler}-16b is the first one that has
been discovered to follow a circumbinary orbit around a system of
two main-sequence stars, a great amount of theoretical and
numerical-experimental work was accomplished in the past three
decades on the hypothetical planetary dynamics in double stellar
systems. A short but comprehensive review of basic theoretical
findings on the relevant stability criteria that can be used to
assess the stability of planetary motion in double stellar
systems, including both inner and outer orbits, is given in
\citep{MW06}.

In particular, criteria, that are most widely used now for these
purposes, were derived in a numerical-experimental way by
\cite{HW99}. An important point is that the fractal structure of
stability/instability borders (in the spaces of orbital elements),
described by these criteria, is averaged out: the borders are
approximated by smooth curves. In our paper we show that the
patterns of the border structure are important for understanding
the current and former dynamical states of {\it Kepler}-16b. Note
that the ``ragged'' structure of the borders in the case of inner
(circumstellar) orbits was analytically revealed by \cite{MW06}.

Recently \cite{PS12} explored the stability of inner
(circumstellar) and outer (circumbinary) planetary orbits in the
$\alpha$~Cen A--B system. Our general approach here is analogous
to that we used in \citep{PS12}. It is as follows. We integrate
test planetary orbits on a fine grid of the starting values of the
orbital pericentric distance and eccentricity, fixing other
orbital elements for a particular epoch. To assess the stability
of the planetary motion, we use two stability criteria. The first
criterion is the value of the maximum Lyapunov exponent. The
second criterion is the ``escape-collision'' one: the orbit is
stable if the planet does not escape (passing to hyperbolic orbit)
from the system, or does not encounter with any of the central
stars.

Resonances in Hamiltonian systems often appear in multiplets;
overlapping of resonances in the multiplets leads to chaos
\citep{C59,C79}. In this paper we show how the {\it Kepler}-16
dynamics serves as a particular illustration to this general rule.
In Section~\ref{sec_sd}, we construct the stability diagrams in
the ``pericentric distance~--- eccentricity'' plane. The location
of the planet {\it Kepler}-16b, which turns out to be situated
just at the edge of chaos domain, in a particular resonance cell,
is analyzed in Section~\ref{sec_ecd}. How these findings relate to
modern scenarios of planet formation is discussed in
Section~\ref{sec_rpfs}. In Section~\ref{sec_rd}, we present a
straightforward dynamical explanation for the fact that the
resonance cells neighboring to that occupied by {\it Kepler}-16b
do not harbor any planet. Section~\ref{sec_asrc} is devoted to a
brief stability analysis of the circumbinary systems {\it
Kepler}-34 and {\it Kepler}-35; planets in these systems also turn
out to be safe in resonance cells. Our basic conclusions are
formulated in Section~\ref{sec_concl}.

\section{Stability diagrams}
\label{sec_sd}

We adopt the values of the orbital parameters of the double star
{\it Kepler}-16 A--B as given in \citep{D11}, and compute the outer
planetary orbits in the planar elliptic restricted three-body
problem. At the initial moment of time, the relative location of
the three bodies is that determined in \citep{D11} for a particular
reference epoch. We vary the planet orbital eccentricity $e$ and
pericentric distance $q = a(1-e)$ (where $a$ is the semimajor axis
of the orbit) and integrate test planetary orbits on a fine grid
of the starting values of $q$ and $e$.

To explore the stability of the planetary motion, we use two
stability criteria. The first criterion is the value of the
maximum Lyapunov exponent. The second criterion is the
``escape-collision'' one: the orbit is stable if the distance
between the planet and one of the stars does not become less than
$10^{-3}$~AU or does not exceed $10^3$~AU.

The Lyapunov exponents of a trajectory measure the rate of
exponential divergence of nearby trajectories in the phase space
of motion \citep{LL92}. Let $d(t_0) \ll 1$ be the initial
displacement of a ``shadow'' trajectory from the ``guiding'' one,
and $d(t)$ be the displacement at time $t$. Then the Lyapunov
exponent is defined by the formula

\begin{equation}
L = \lim_{{t \to \infty} \atop {d(t_0) \to 0}} {1 \over {t-t_0}}
\ln{d(t) \over d(t_0)}. \label{LCE}
\end{equation}

\noindent Depending on the direction of the initial displacement
in the phase space, the quantity $L$ of a trajectory of a
Hamiltonian system can attain $2N$ generally different values,
where $N$ is the number of degrees of freedom. These values form
symmetric pairs: for each $L_i > 0$ there exists $L_{i+N} = - L_i
< 0$; $i=1$, \ldots, $N$ \citep{LL92}. Therefore, in practice it
is sufficient to compute solely $N$ exponents $L_i \ge 0$. The set
of all $2N$ exponents is called the spectrum of Lyapunov
exponents, or, the Lyapunov spectrum. A non-zero value of the
maximum (in the spectrum) Lyapunov exponent indicates chaotic
character of the motion, whereas zero value indicates that the
motion is regular, i.e., quasiperiodic or periodic
\citep{C79,LL92}. On an everywhere dense set of starting data for
shadow trajectories, the Lyapunov exponent attains a single (the
maximum) value, --- the maximum Lyapunov exponent, which we denote
as $L$ in what follows. The inverse of this quantity,
$T_\mathrm{L} \equiv L^{-1}$, is the so-called Lyapunov time. It
represents the characteristic time of predictable dynamics.

For computing the Lyapunov spectra we use the algorithms and
software developed in \citep{SK02,KS03,KS05} on the basis of the
HQRB method by \cite{BUP97}. This method is based on the QR
decomposition of the tangent map matrix using the Householder
transformation. The trajectories are integrated using the
integrator by \cite{HNW87}. It is an explicit 8th order
Runge--Kutta method due to Dormand and Prince, with the step size
control.

For separating the regular and chaotic orbits of a dynamical
system, a statistical method was proposed in
\citep{MS98,S02c,SM03}. It consists of four steps. ({\it i}) On a
representative set of initial data, two differential distributions
(histograms) of the orbits in the computed value of $\log_{10}
T_\mathrm{L}$ are constructed, using two different time intervals
for the integration. ({\it ii}) If the phase space of motion is
divided \citep{C79}, each of these distributions has at least two
peaks. The peak that shifts (moves in the positive direction of
the horizontal axis), when the integration time interval is
increased, corresponds to the regular orbits. The fixed peak (or
peaks) corresponds (correspond) to the chaotic orbits. ({\it iii})
The value of $\log_{10} T_\mathrm{L}$ at the histogram minimum
between the peaks is identified, giving a numerical criterion for
separating the regular and chaotic orbits. ({\it iv}) The obtained
criterion can be used in any further computations to separate the
regular and chaotic orbits on much finer initial data grids and,
rather often, on smaller time intervals of integration.

%Figures 1, 2, 3 here.

We consider two representative sets of initial data. They are
formed as fine grids in rectangular areas in the ($q$,~$e$) plane,
namely, in the areas ($0.5 \leq q \leq 1.0$, $0 \leq e \leq 0.9$)
and ($0.6 \leq q \leq 0.8$, $0 \leq e \leq 0.2$). The second area,
which is smaller, is explored with a higher resolution in $q$ and
$e$, i.e., the ($q$,~$e$) grid is finer.

In Fig.~\ref{distr}, the distributions (histograms) $f$ of
planetary orbits in the computed value of $\log_{10} T_\mathrm{L}$
for the first (large) area are shown for two choices of the
computation time: that computed on the time interval of $10^3$~yr
is shown in red, and that computed on the time interval of
$10^4$~yr is shown in blue. The quantity $f$ is defined as the
number of orbits with $\log_{10} T_\mathrm{L}$ in the interval
$(\log_{10} T_\mathrm{L}$, $\log_{10} T_\mathrm{L} + \Delta
\log_{10} T_\mathrm{L})$, normalized by the total number of orbits
in the set; $\Delta \log_{10} T_\mathrm{L}$ is set equal to
$0.02$. As follows from the histograms in Fig.~\ref{distr}, the
threshold (separating chaos and order) value of $\log_{10}
T_\mathrm{L}$ can be set equal to $2.5$.

The resulting stability diagrams are shown in Figs.~\ref{st_diag}
and \ref{st_diag_hr}. Fig.~\ref{st_diag}$a$ is constructed using
the Lyapunov exponent criterion, and Fig.~\ref{st_diag}$b$ using
the Lyapunov exponent plus escape-collision criteria. In
Fig.~\ref{st_diag_hr}, the stability diagrams are shown in a
higher resolution in $q$ and $e$.

A direct inspection of Figs.~\ref{st_diag}$b$ and
\ref{st_diag_hr}$b$ demonstrates that the Lyapunov exponent
criterion provides a more clear-cut picture of the chaos-order
borders, in comparison with using the escape-collision criterion
at the same time interval of integration. What is more, the
escape-collision criterion fails to identify orbits with high
values of the pericentric distance and eccentricity, at least at
the specified intervals of computation time. The chaos-order
borders, revealed by the Lyapunov exponent criterion, apparently
possess fractal structure conditioned by the orbital resonances
(discussed below).

\section{{\it Kepler}-16b: at the edge of chaos domain}
\label{sec_ecd}

The location of the planet {\it Kepler}-16b is shown in
Figs.~\ref{st_diag} and \ref{st_diag_hr} by a dot. As it is clear
from the diagrams, the planet turns out to be located almost at
the edge of the chaos domain, in a hazardous vicinity to it~---
just between the instability ``teeth''. A direct linear
extrapolation of these ``teeth'' to the $e=0$ axis shows that they
correspond to integer resonances between the orbital periods of
the planet and the central binary. The two teeth surrounding {\it
Kepler}-16b correspond to the 5/1 and 6/1 resonances: at these
resonances, the orbital periods of the planet and the central
binary are in the ratios 5/1 and 6/1. The smaller teeth centered
between the ``integer'' teeth correspond to half-integer
resonances. The tooth almost pointing at {\it Kepler}-16b
corresponds to the 11/2 resonance.

Why there is no instability in this resonance, at the location of
{\it Kepler}-16b, whereas the neighboring ``integer'' teeth
extend down to the $e=0$ axis?

Chaotic behaviour, which is often present in the dynamics of
celestial bodies, is usually due to interaction of resonances (as
in any Hamiltonian system, see \citealt{C79}). How to distinguish
between resonant and non-resonant motions? In fact, the observable
commensurability between the orbital frequencies never happens to
be ideally exact, at least due to observational errors. To solve
this problem, a ``resonant argument'' (synonymously, ``resonant
phase'' or ``critical argument'') is introduced. It is a linear
combination of some angular variables of a system under
consideration. In our notations, the resonant argument for an
outer resonance (a resonance between a central binary and a
circumbinary particle) is defined by the formula
\citep{MD99,Morbi02}:

\begin{equation}
  \sigma = (k+q)\lambda_\mathrm{p} - k\lambda_\mathrm{s} - l\varpi_\mathrm{p},
  \label{res_def1}
\end{equation}

\noindent where $\lambda_\mathrm{s}$ and $\lambda_\mathrm{p}$ are
the mean longitudes of a star (in the central binary) and a
particle, respectively; $\varpi_\mathrm{p}$ is the longitude of
pericenter of the particle; $k$, $q$ and $l$ are integers; $q$ is
the so-called {\it resonance order}. (Note that in
Eq.~(\ref{res_def1}) the longitude of pericenter of the star is
ignored, because it is practically constant.) In outer resonance,
the ratio of orbital periods of a planet and the central binary is
equal to $(k+q)/k$.

Usually a mean motion resonance splits in a multiplet of
subresonances, corresponding to a sequence of values of $l$. This
phenomenon is due to precession of particle's orbit pericenter,
often taking place when the motion is perturbed
\citep{HM96,MH97,Morbi02}. Here the perturbation is strong,
because the mass parameter $\mu = M_2/(M_1 + M_2)$ (where $M_1 >
M_2$ are the stellar masses) is large: it is equal to $0.227$
\citep{D11}. Hence the precession of the planetary orbit is
strong.

The theory of resonances splitting in multiplets was developed in
\citep{HM96,MH97} for the high-order inner mean motion resonances
in the dynamics of asteroids. According to this theory, inner mean
motion resonance $k/(k+q)$ splits in a cluster of $q+1$
subresonances with $l = 0, 1, \ldots, q$. Analogously, the outer
mean motion resonances also split.

The coefficients of the subresonant terms in the expansion of the
perturbing function are proportional to the eccentricities of the
particle and perturber in some powers depending on $q$; in
particular, the coefficients of the first and last subresonant
terms in the multiplet are proportional, respectively, to the
eccentricities of the perturber and particle in the power equal to
$q$ \citep{HM96,MH97}. In the pendulum model of subresonance, its
width (characterizing the ``subresonance strength'') in the phase
space is proportional to the square root of this coefficient
\citep{C79,HM96}. Thus the value of $q$ governs this basic
property of the resonant motion.

Consider two neighboring outer integer resonances $(q+1)/1$ and
$(q+2)/1$, which have orders $q$ and $q+1$, respectively. The
half-integer resonance between them is $(2q+3)/2$, and its order
is $2q+1$. Thus, for a high-order half-integer resonance, the
power-law indices in the subresonant term coefficients are much
greater than the indices for the neighboring integer resonances;
consequently, the strengths of subresonances are much less and
their interaction is much weaker. On increasing $e$, they start to
overlap much later than in the neighboring integer cases. This
explains why the 11/2 resonance at the eccentricity of {\it
Kepler}-16b is stable, whereas the neighboring integer resonances
5/1 and 6/1 are unstable. Concluding, {\it Kepler}-16b survives
because it is close to the half-integer 11/2 orbital resonance
with the central binary. The planet is safe inside a resonance
cell bounded by the unstable 5/1 and 6/1 resonances.

In the Solar system, this phenomenon is analogous to the survival
of Pluto and Plutinos. They are in the 3/2 outer orbital resonance
with Neptune; thus the order of the occupied high-integer
resonance is much smaller than in the {\it Kepler}-16 system. This
is because the mass parameter $\mu$ in the case of Neptune and the
Sun is much smaller and therefore the chaos border radically
shifts to smaller values of the semimajor axis (in units of the
``central binary'' size).

The analogy with the case of resonant trans-Neptunian objects is
striking. According to \cite{G12}, the population of TNOs in the
next half-integer resonance (5/2) with Neptune is estimated to be
as large as in the 3/2 resonance, whereas other (non-half-integer)
resonant populations are radically smaller. The most distant known
resonant TNO is in the 27/4 resonance with Neptune \citep{G12};
note that this is farther than the 11/2 resonance. If an object in
the 11/2 resonance with Neptune were once discovered, it could be
called a ``{\it Kepler}-16b of the Solar system''.

\section{Resonances and the planet formation scenarios}
\label{sec_rpfs}

Although from the viewpoint of the discussed above
``trans-Neptunian analogy'' the current planetary architecture of
{\it Kepler}-16 does not seem peculiar, it looks so in the light
of modern theories of planet formation in circumbinary systems.
Indeed, it might seem from the diagram in Fig.~\ref{st_diag_hr}
that no radial migration was possible for {\it Kepler}-16b since
its formation epoch, because otherwise it would cross the
instability ``teeth'' and thus would be removed. On the other
hand, in situ formation of {\it Kepler}-16b is a theoretical
challenge \citep{M12,P12}. Let us consider this contradiction in
more detail.

First of all, the presence of zones of instability on the
migration path does not necessarily mean catastrophic
consequences. From another field of celestial mechanics, one may
recall that the satellites of planets in the Solar system are
known to be slowly despun, in the process of tidal evolution,
until they reach the well known 1:1 synchronous spin-orbit
resonance, and in the course of despinning a number of chaotic
layers in the phase space of motion is crossed \citep{W87}; the
broadest layer is at the separatrix of the synchronous resonance.
Nevertheless, all tidally-evolved satellites (with a possible
exception of no more than three satellites, see
\citealt{KS05,MS10}) are lucky to reside in the ``final'' stable
synchronous resonance, as the Moon does. Evidently, most of the
satellites were able to cross the chaotic layers without being
caught in chaos forever. The reason is that the timescales for
development of gross instability are usually too long in
comparison with the timescales for crossing the chaotic zones
\citep{W87,KS05}. In the considered case of planetary migration
the timescale for development of gross instability might be as
well too long in comparison with the timescale for crossing the
chaotic zone. More long-term numerical experiments are needed to
verify this possibility.

Concerning possible scenarios for formation of circumbinary
planets, modern theories and simulations favor, within the planet
accretion framework, the following one: the planetary core forms
further out (in an accretion-friendly region) in the
protoplanetary disk and then migrates inward until the migration
is stalled at the border of the disk inner cavity formed by the
central binary \citep{PN07,M12,P12}. The cavity might be
comparable in size to the chaos region (explored above) for test
particles. \cite{P12} estimate the final locations of {\it
Kepler}-16b, 34b and 35b to be close to the truncation radii of
the gas disks. Though in situ formation of {\it Kepler}-16b is
still not ruled out \citep{M12}, it is less likely owing to
hostile conditions (in particular, high encounter velocities of
planetesimals and low planetesimal density) for planetesimal
accretion \citep{M12,P12}. One may speculate that a possible
scenario could be that the inward migration of {\it Kepler}-16b is
stopped at its currently observed location due to capture in the
11/2 resonance.

The current location of {\it Kepler}-16b close to the center of a
resonance cell should be taken into account and explained when
constructing formation scenarios for the planet. In a general
framework of circumbinary planet formation theories, an important
role of orbital resonances was mentioned and considered in
\citep{MN04,PN07,PN08}. In particular, \cite{MN04} pointed out
that the pumped eccentricities of planetesimals, in function of
the semimajor axis, show ``interesting behavior such as somewhat
resonant features'' (see fig.~1 in \citealt{MN04}). Such
resonances as 5/1 affect the formation and orbital evolution of
giant Saturn-mass planets embedded in a circumbinary disc, as the
results of hydrodynamic simulations by \cite{PN07,PN08} show.

Finally, it is worth noting that at least two astrophysical
processes are known that can lead to material occupation of
high-order outer resonances in relevant gravitational systems. As
it is well known, in a dynamical (e.g., any gravitational or
planetary) system whose parameters are slowly (adiabatically)
varying, capture in resonances can occur; in particular, the above
mentioned Plutinos are believed to have been trapped in the 3/2
resonance with Neptune in the Kuiper belt due to the outward
migration of Neptune; generally, the outward migration of the
perturbing body (planet) in a planetary system can lead to capture
of particles in outer mean motion resonances; see \citep{Q06} and
references therein. Another well-known astrophysical process, that
leads to material occupation of outer resonances, takes place in
dusty debris disks around stars with planets: when, due to
dissipational forces, dust spirals inward, it can be efficiently
captured in resonances; see \citep{DM05,Q06} and references
therein. \cite{DM05} accomplished simulations of the debris disk
evolution in the Fomalhaut system using a planet with mass equal
to doubled mass of Jupiter; striking examples of dense occupation
of high-order outer resonances were demonstrated: see, e.g.,
Fig.~14 in \citep{DM05}, where integer (such as 4/1 and 5/1) and
half-integer (such as 5/2, 7/2, and 9/2) resonances dominate or
are prominent in the ``semimajor axis~--- resonance occupation''
diagram.

\section{{\it Kepler}-16b as a ``resonant destroyer''}
\label{sec_rd}

Then, another question arises: why is there only one resonance
cell that is ``occupied''? Why there is not a lot (or several)
planets in resonance cells like bees in a honeycomb? At least for
the cells neighboring to {\it Kepler}-16b the answer is
straightforward. Again, it concerns resonances (and their
interaction), though different from those considered above. These
are the first-order orbital resonances $(k+1)/k$ of test particles
with the planet. On increasing $k$, these resonances start to
overlap at some critical $k$, because their widths do not decrease
fast enough. On such grounds \cite{W80} inferred that in the case
of small eccentricity ($e < 0.15$) of the particle orbit the
critical $k$ is given by

\begin{equation}
k_\mathrm{overlap} \approx 0.51 \mu_\mathrm{p}^{-2/7},
\end{equation}

\noindent where $\mu_\mathrm{p} = M_\mathrm{p} / (M_\mathrm{s} +
M_\mathrm{p})$ is the mass parameter, and $M_\mathrm{s}$ and
$M_\mathrm{p}$ are the masses of the star and the planet,
respectively. Using the third Kepler law, one finds that $k =
k_\mathrm{overlap}$ corresponds to the chaotic zone width

\begin{equation}
\Delta a_\mathrm{overlap} \approx 1.3
\mu_\mathrm{p}^{2/7}a_\mathrm{p}, \label{em_27}
\end{equation}

\noindent where $a_\mathrm{p}$ is the semimajor axis of the planet
\citep{DQT89,MD99}. The particles with $a$ within the interval
$a_\mathrm{p} \pm \Delta a_\mathrm{overlap}$ move chaotically. Due
to encounters with the planet they escape from this region sooner
or later; in such a way a particle-free zone around the planet
orbit is formed.

Using Eq.~(\ref{em_27}) and data on $a_\mathrm{p}$ and masses from
\citep{D11}, and setting $M_\mathrm{s} = M_1 + M_2$, one finds for
{\it Kepler}-16b: $\mu_\mathrm{p} = 3.56 \cdot 10^{-4}$ and
$\Delta a_\mathrm{overlap} \approx 0.095$~AU. Therefore, at least
the two neighboring resonance cells, namely those centered at the
9/2 and 13/2 resonances, are purged by the planet residing in the
11/2 resonance cell, because they are within the $\Delta a \approx
0.1$~AU distance (see Figs.~\ref{st_diag} and \ref{st_diag_hr}).
Thus {\it Kepler}-16b can be called not only a ``resonant
survivor'' but a ``resonant destroyer'' as well.

\section{{\it Kepler}-34b and {\it Kepler}-35b: also safe in resonance cells}
\label{sec_asrc}

Two new circumbinary planetary systems were discovered recently,
{\it Kepler}-34 and {\it Kepler}-35 \citep{W12}. Let us see how
the stability diagrams look like for the planets in these systems.
Both systems are single-planet. We take the necessary data on the
masses and orbital elements from \citep{W12} and perform
computations analogous to those described above in
Section~\ref{sec_sd} for the {\it Kepler}-16 case.

%Figures 4, 5 here.

In comparison with {\it Kepler}-16, the histograms of the orbits
with respect to the computed Lyapunov time value can be computed
on smaller time intervals, because the binaries' periods in the
{\it Kepler}-34 and {\it Kepler}-35 systems are much smaller. For
speed of computation, we have chosen the time intervals equal to
$10^2$ and $10^3$~yr, instead of $10^3$ and $10^4$~yr in the case
of {\it Kepler}-16. Analysis of the constructed histograms leads
to the threshold (separating chaos and order) value of $\log_{10}
T_\mathrm{L}$ equal to $1.5$ for both {\it Kepler}-34 and {\it
Kepler}-35. Apparently, here the threshold $T_\mathrm{L}$ is 10
times less than in the case of {\it Kepler}-16, because the time
intervals of computation are 10 times less.

In Figs.~\ref{st_diag34} and \ref{st_diag35}, the stability
diagrams computed for {\it Kepler}-34b and {\it Kepler}-35b are
shown. One can see that qualitatively these diagrams are analogous
to those computed for {\it Kepler}-16b (i.e., to
Figs.~\ref{st_diag}b and \ref{st_diag_hr}b): the planets are
inside resonance cells; these cells are bounded by the ``teeth''
of unstable resonances. Though the planets are in a dangerous
vicinity to the chaos domain, they are both safe inside the
resonance cells. As follows from data in \citep{W12}, the ratios
of orbital periods of the planet and the stellar binary in each of
the systems are equal to 10.4 and 6.34, respectively. Judging from
these values, the cells might be centered, respectively, at the
half-integer 21/2 and 13/2 resonances with the central binaries;
though in the first case, due to the very high order of the
resonance, the situation is not certain.

While this paper was in the reviewing process, two more
circumbinary planetary systems have been discovered: {\it
Kepler}-38 \citep{O12a} and {\it Kepler}-47 \citep{O12b}.
Moreover, the latter one is a multi-planet system, hosting at
least two planets, planet $c$ moving in a much larger orbit than
planet $b$.

The ratios of orbital periods of the planet $b$ and the stellar
binary, as follows from data in \citep{O12a,O12b}, are equal to
5.62 and 6.65 in the {\it Kepler}-38 and {\it Kepler}-47 systems,
respectively. Thus the planetary dynamical states in these systems
might be similar to those in the {\it Kepler}-16 and {\it
Kepler}-35 systems (where the period ratios are 5.57 and 6.34,
respectively), i.e., the planets might occupy the same resonance
cells, centered at the half-integer 11/2 and 13/2 resonances,
respectively.

\section{Conclusions}
\label{sec_concl}

Our basic conclusions are as following.

({\it i}) The minimum outer border of the chaotic domain in the
stability diagram for the {\it Kepler}-16 system corresponds to
the semimajor axis $0.60$~AU at the zero eccentricity of a test
planet.

\begin{figure}[ht!]
\center{\includegraphics[width=7cm]{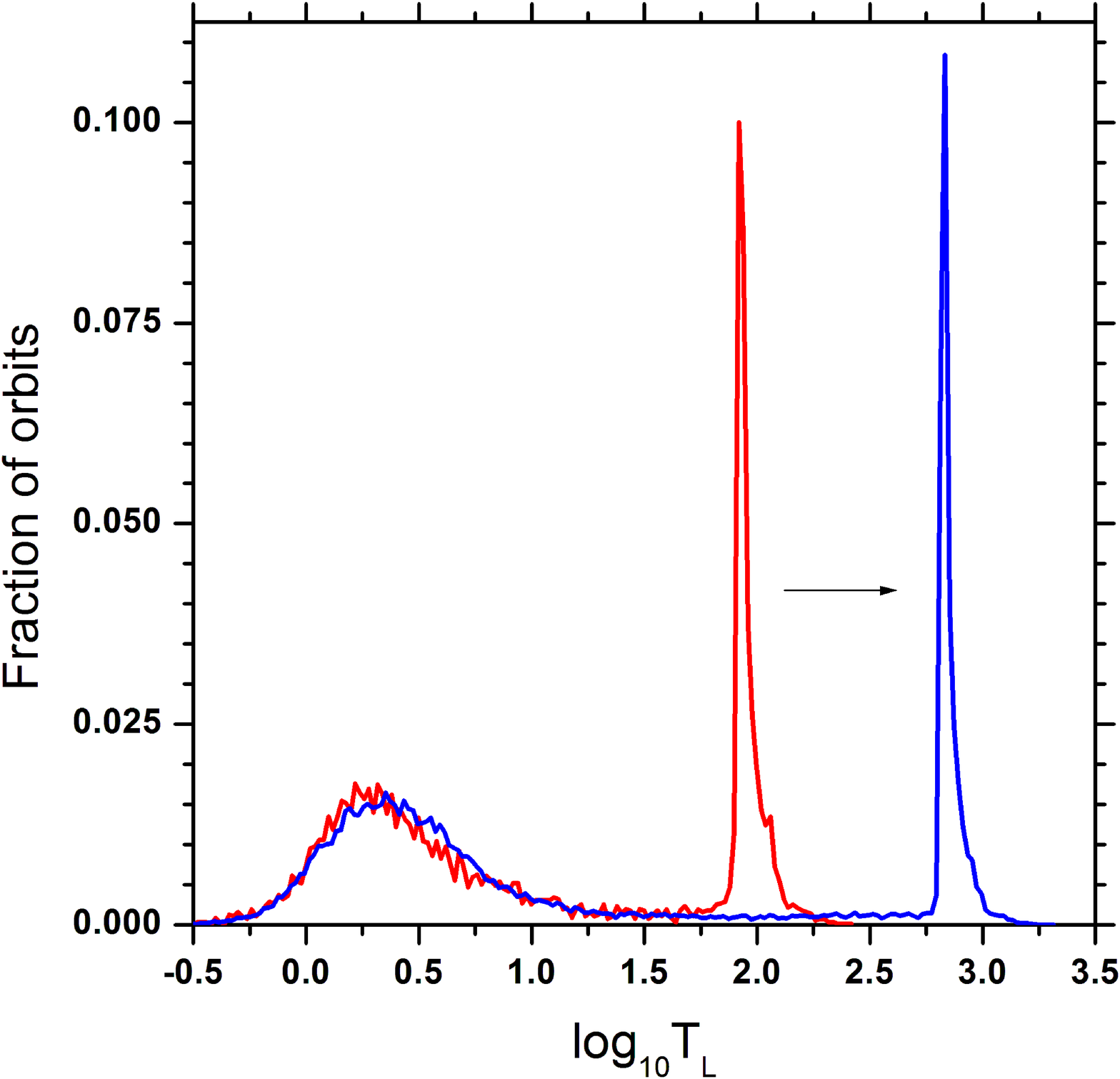}} \caption{The
histograms of circumbinary orbits in the {\it Kepler}-16 system in
the Lyapunov time logarithm. The histogram of the orbits computed
on the time interval of $10^3$~yr is shown in red, and that
computed on the time interval of $10^4$~yr is shown in blue.}
\label{distr}
\end{figure}

\begin{figure}
\begin{center}
\begin{tabular}{ll}
a)~\includegraphics[width=7cm]{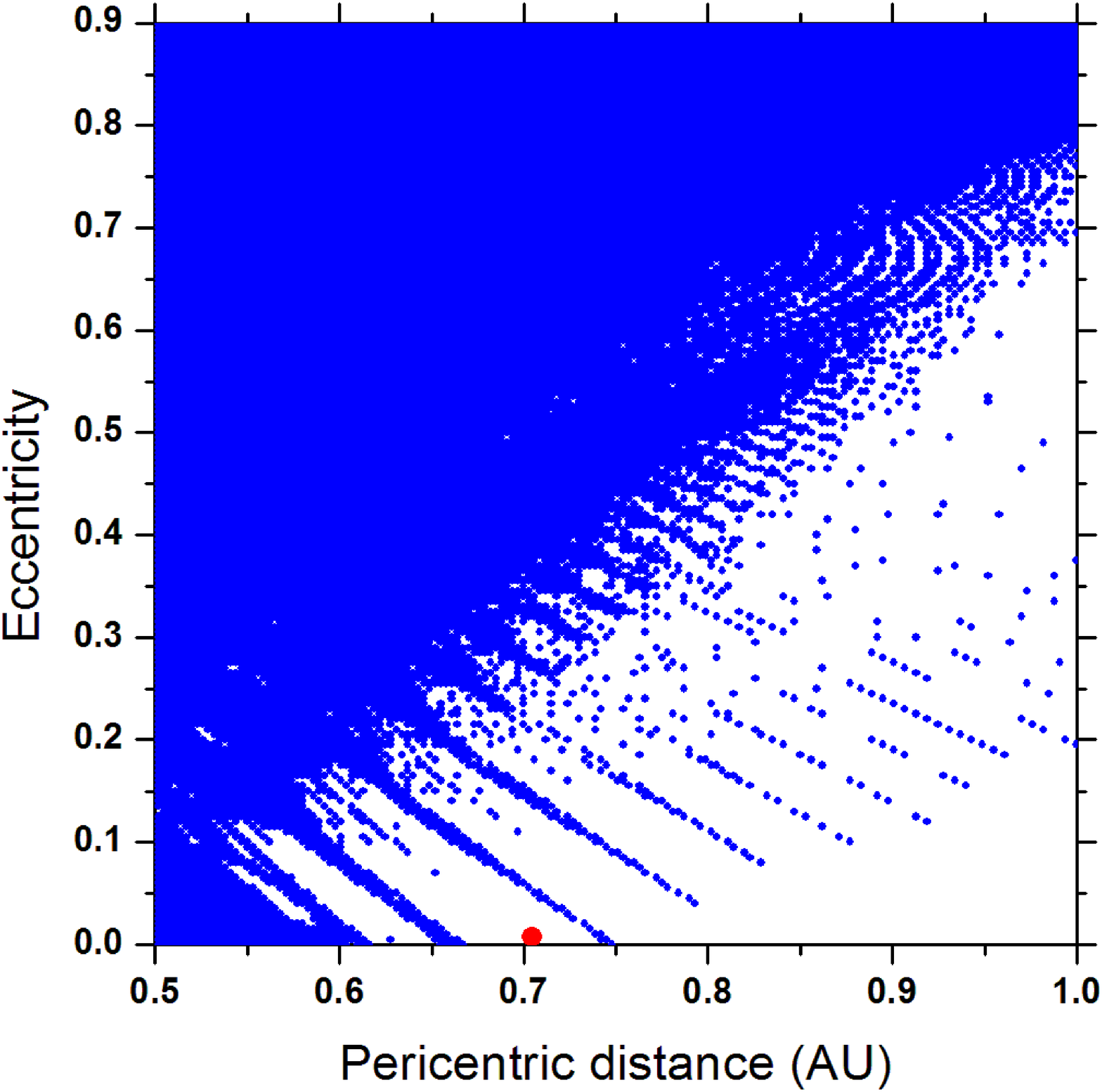} &
b)~\includegraphics[width=7cm]{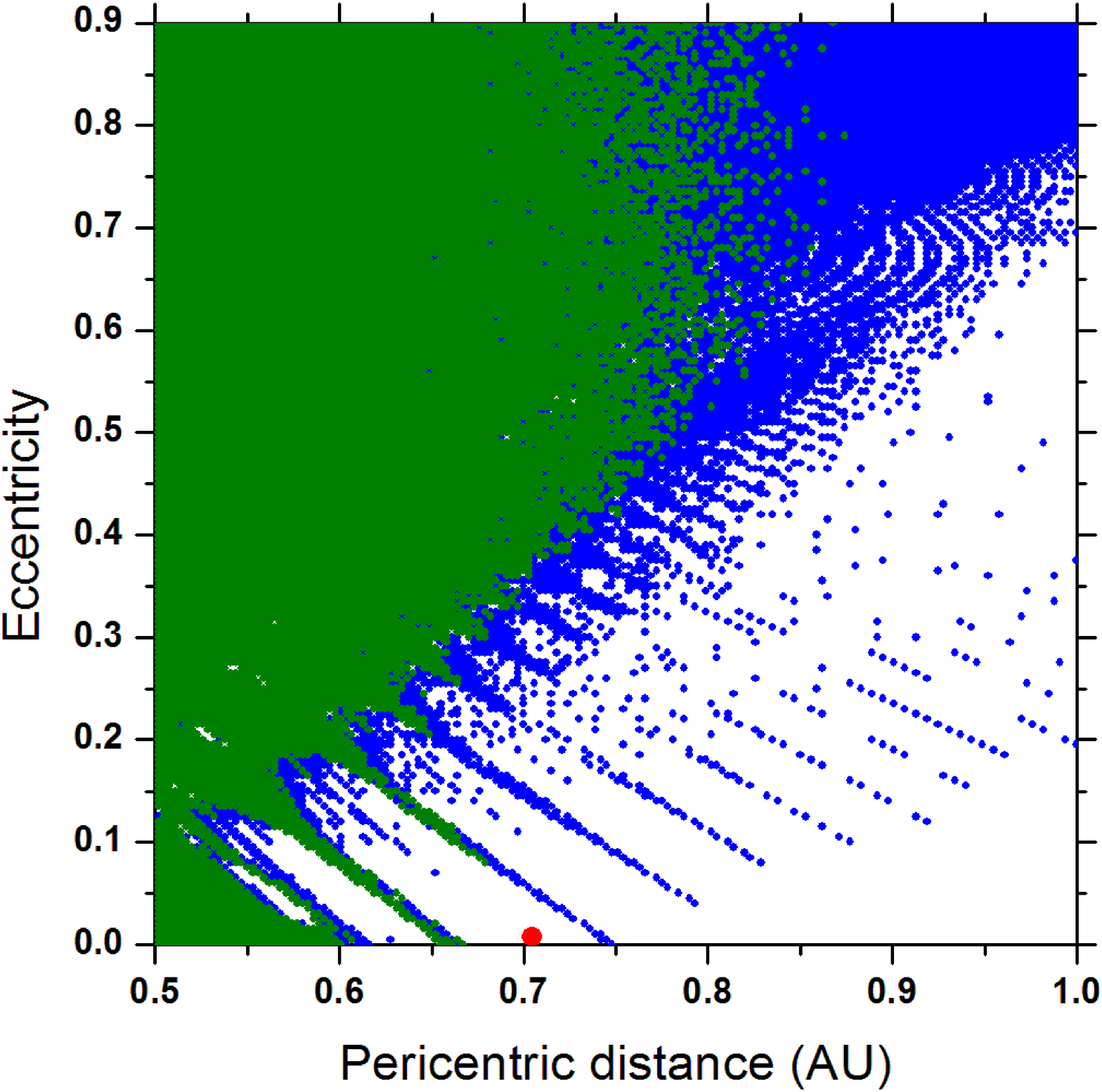}
\end{tabular}
\end{center}
\caption{\small The stability diagrams constructed by the Lyapunov
exponent criterion (a) and by the Lyapunov exponent plus
escape-collision criteria (b). The regular regions are shown in
white, chaotic in blue, and those for orbits with escapes and
collisions (at the adopted time interval) in green. The location
of the {\it Kepler}-16b planet is shown by a dot.}
\label{st_diag}
\end{figure}

({\it ii}) The representative values of the Lyapunov time
(corresponding to the first (left) peak of the distribution in
Fig.~\ref{distr}) for the outer orbits in the chaotic domain are
$\sim 2$~yr.

({\it iii}) The Lyapunov exponent criterion, applied for the
construction of the stability diagrams, provides much better
resolution of the chaos-order borders in the stability diagrams in
comparison with the escape-collision criterion.

({\it iv}) The planet {\it Kepler}-16b turns out to be almost at
the edge of chaos domain, in a hazardous vicinity to it~--- just
between the instability ``teeth'' in the space of orbital
parameters. {\it Kepler}-16b survives, because it is safe inside a
resonance cell bounded by the unstable 5/1 and 6/1 resonances.
What is more, the planet is rather close to the stable
half-integer 11/2 orbital resonance with the central binary.

({\it v}) A straightforward dynamical explanation for the fact
that the resonance cells neighboring to that occupied by {\it
Kepler}-16b are vacant (do not harbor any planet) is that they are
``purged'' by {\it Kepler}-16b, due to overlap of first-order
resonances with the planet.

({\it vi}) In the Solar system, the survival of {\it Kepler}-16b
can be called analogous to the survival of Pluto and Plutinos,
which are in the outer 3/2 orbital resonance with Neptune, and
other objects in outer half-integer higher-order orbital
resonances with Neptune. The minimum order of the ``occupied''
half-integer resonance grows with the mass parameter $\mu$ of the
perturbing binary (in the case of the Solar system, the relevant
``binary'' is the Sun and Neptune), because increasing $\mu$
shifts the stability border outwards.

\begin{figure}
\begin{center}
\begin{tabular}{ll}
a)~\includegraphics[width=7cm]{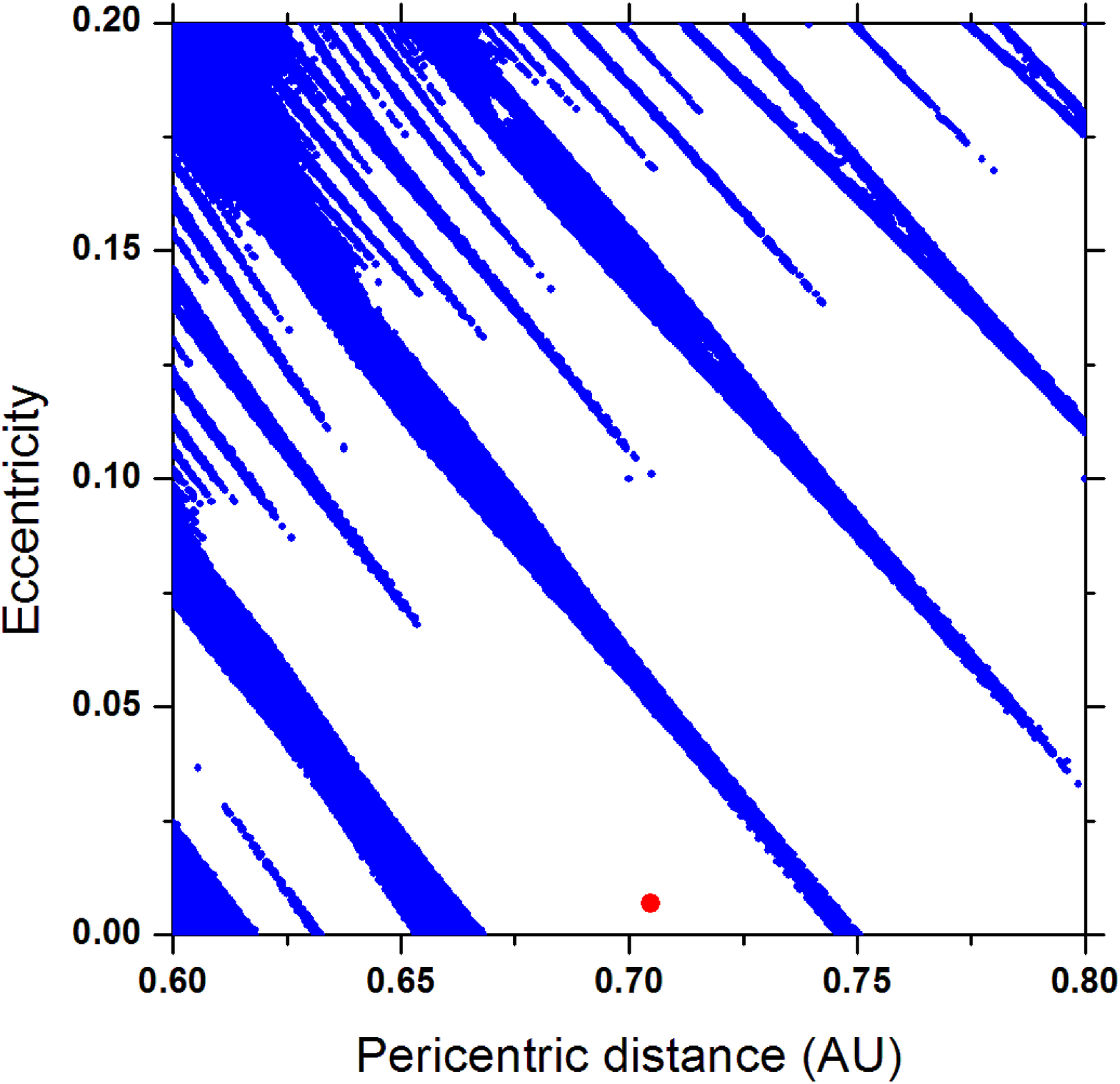} &
b)~\includegraphics[width=7cm]{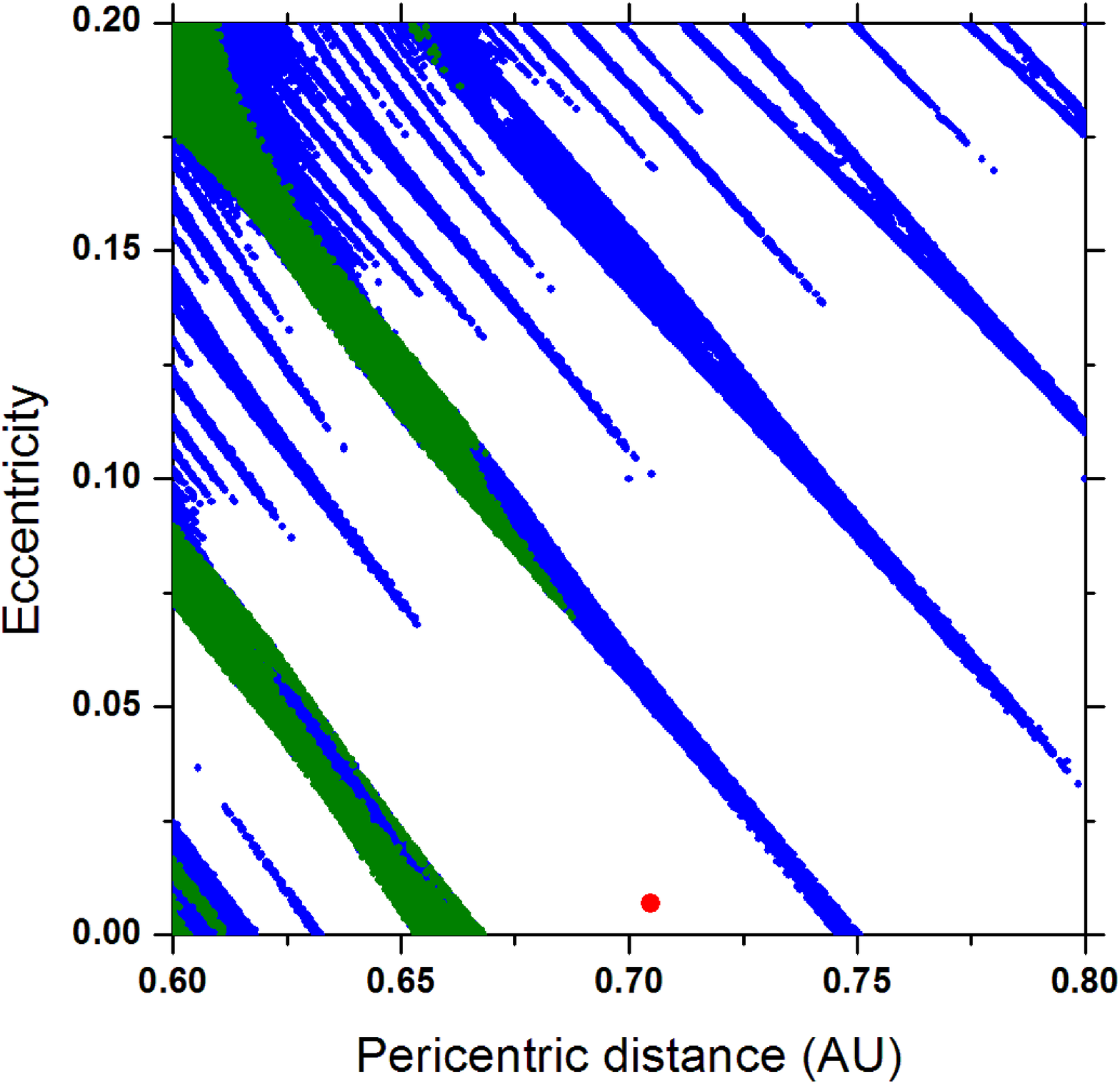}
\end{tabular}
\end{center}
\caption{\small Zooms of the stability diagrams in
Fig.~\protect\ref{st_diag}, in a higher resolution.}
\label{st_diag_hr}
\end{figure}

\begin{figure}
\begin{center}
\begin{tabular}{ll}
a)~\includegraphics[width=7cm]{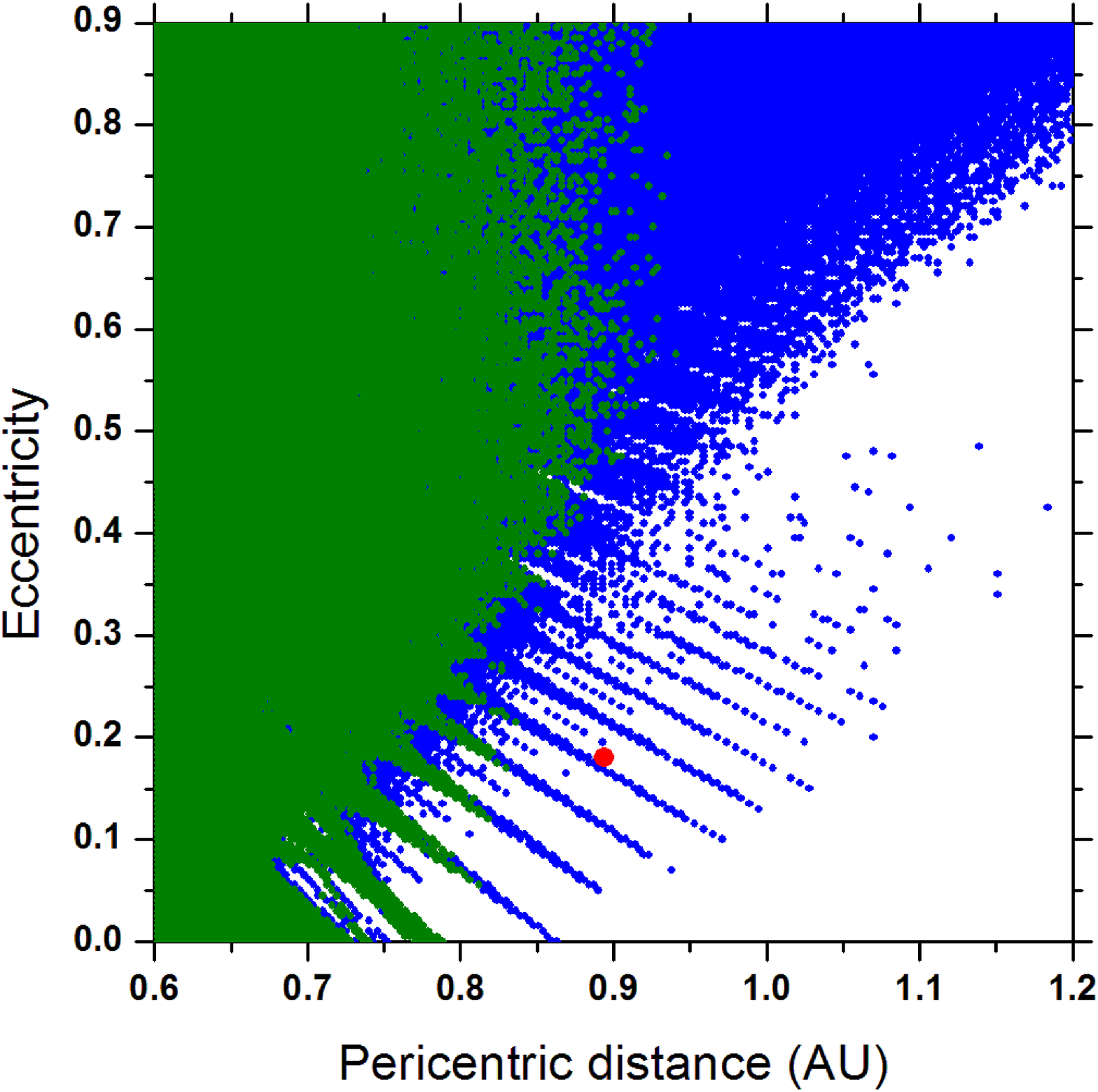} &
b)~\includegraphics[width=7.13cm]{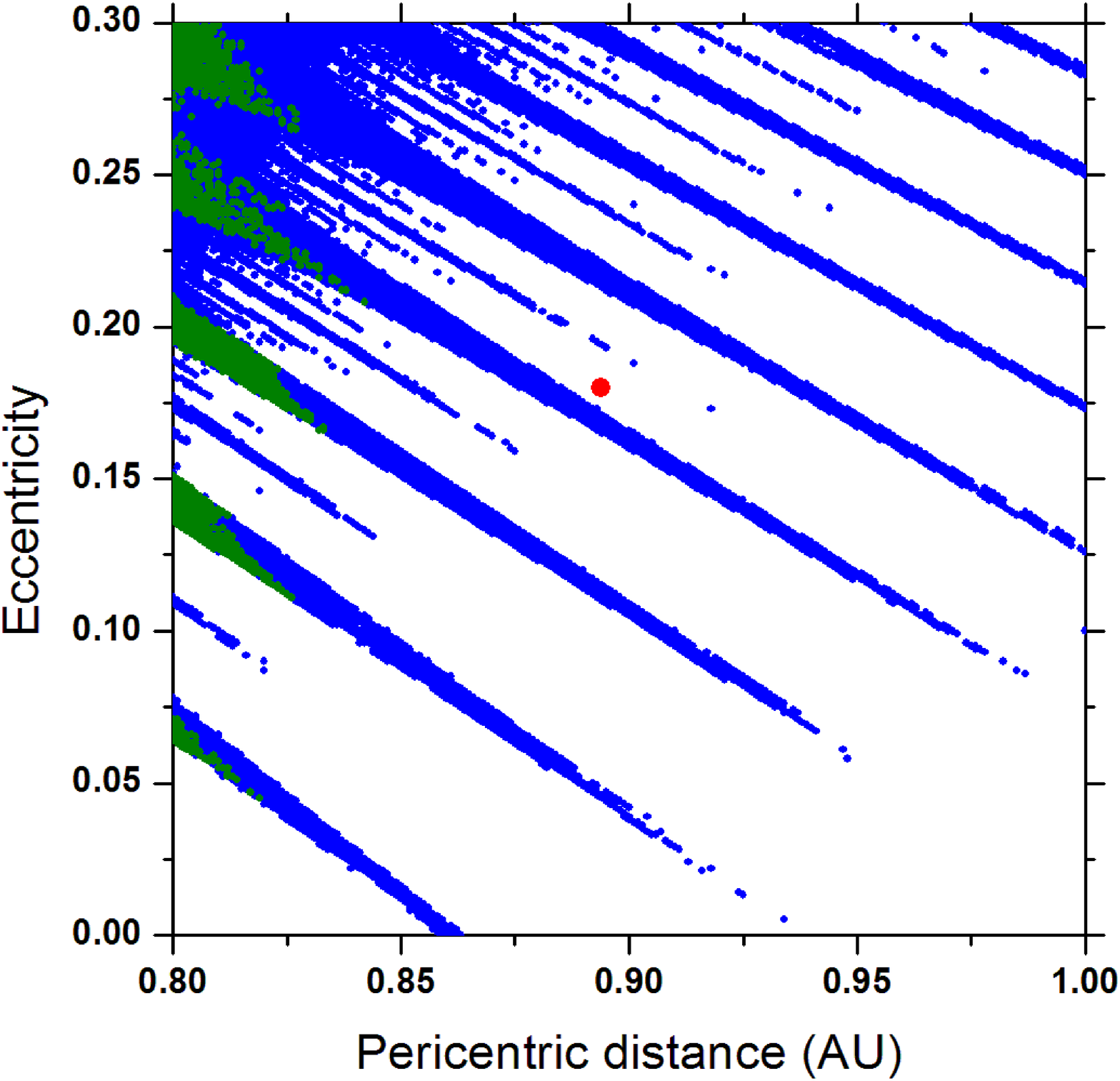}
\end{tabular}
\end{center}
\caption{\small The same as in Figs.~\protect\ref{st_diag}b and
\protect\ref{st_diag_hr}b, but for {\it Kepler}-34b.}
\label{st_diag34}
\end{figure}

\begin{figure}
\begin{center}
\begin{tabular}{ll}
a)~\includegraphics[width=7cm]{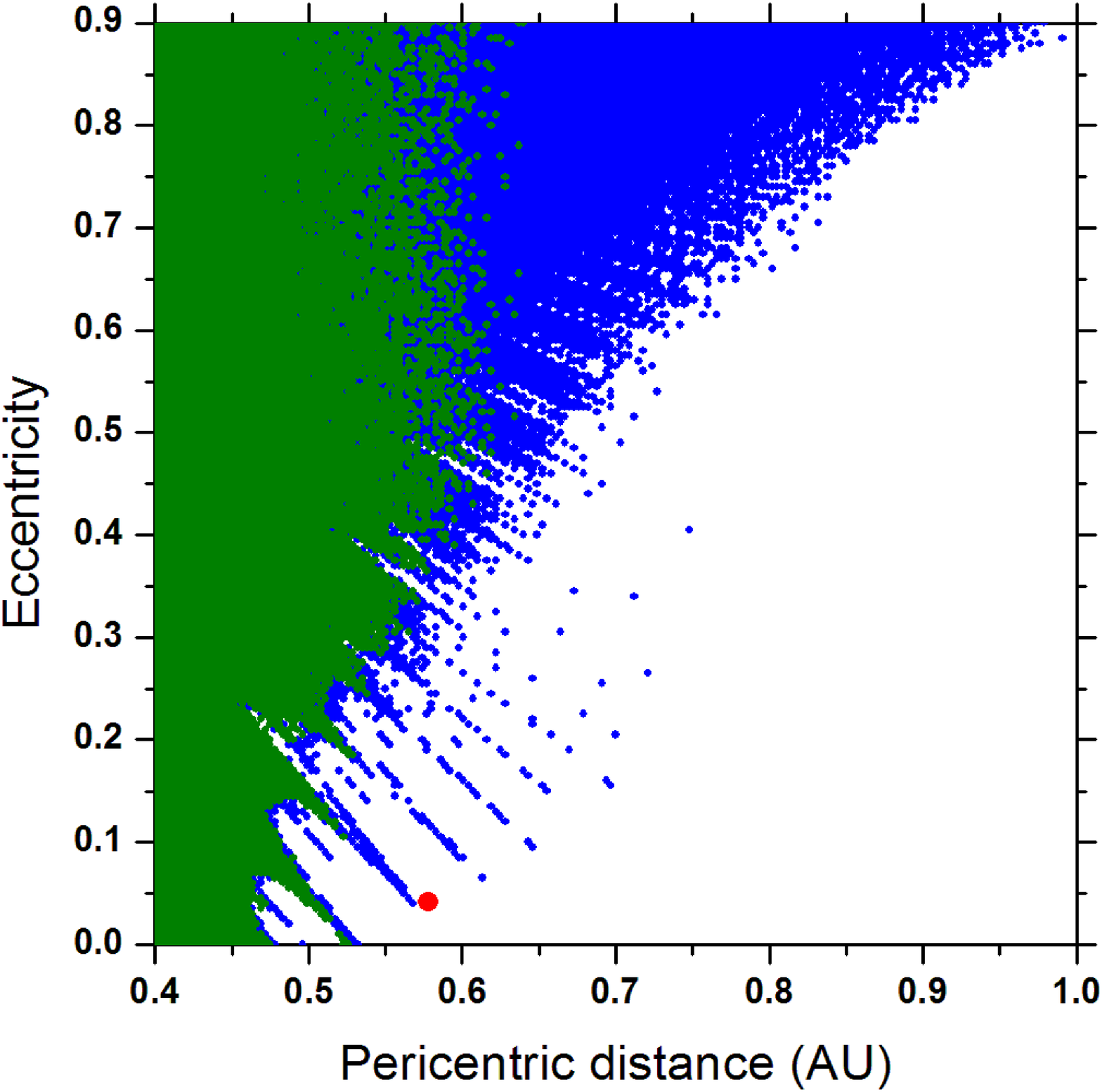} &
b)~\includegraphics[width=7.13cm]{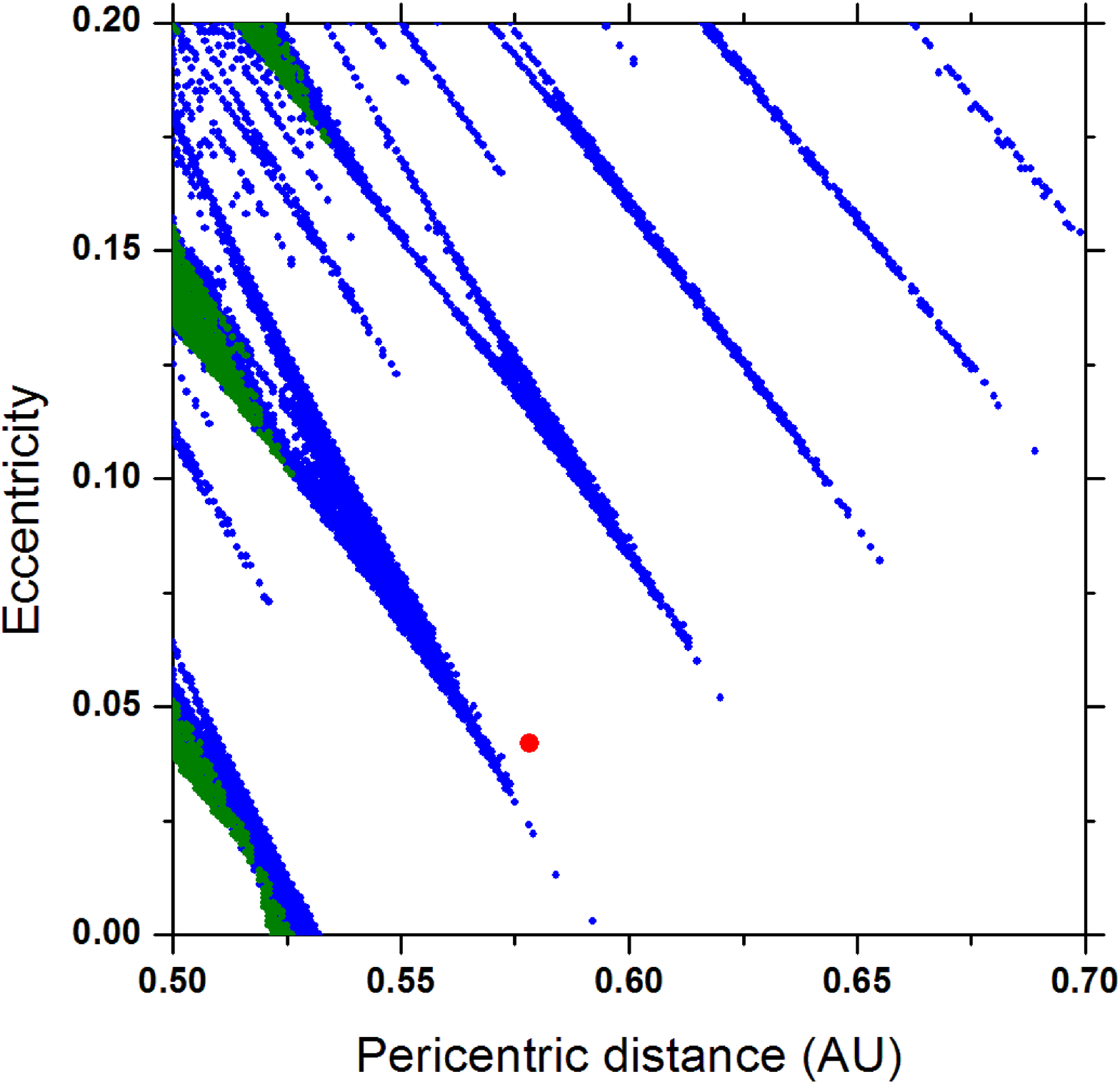}
\end{tabular}
\end{center}
\caption{\small The same as in Figs.~\protect\ref{st_diag}b and
\protect\ref{st_diag_hr}b, but for {\it Kepler}-35b.}
\label{st_diag35}
\end{figure}

({\it vii}) From the diagram in Fig.~\ref{st_diag_hr}, it might
seem that no radial migration was possible for {\it Kepler}-16b
since its formation, because otherwise it would cross the
instability ``teeth'' and thus would be removed. However, the
timescale for development of gross instability might be too long
in comparison with the timescale for crossing the chaotic zone.
More long-term numerical experiments are needed to verify this
possibility. On the other hand, in situ formation of {\it
Kepler}-16b is a theoretical challenge \citep{M12,P12}. The
current location of {\it Kepler}-16b close to the center of a
resonance cell should be taken into account and explained when
constructing formation scenarios for the planet.

({\it viii}) The newly discovered planets {\it Kepler}-34b and
{\it Kepler}-35b are also safe inside resonance cells at the chaos
border.

\bigskip

The authors are grateful to the referee for useful remarks and
comments. This work was supported in part by the Russian
Foundation for Basic Research (project No.\ 10-02-00383) and by
the Programmes of Fundamental Research of the Russian Academy of
Sciences ``Fundamental Problems in Nonlinear Dynamics'' and
``Fundamental Problems of the Solar System Studies and
Exploration''. The computations were partially carried out at the
St.~Petersburg Branch of the Joint Supercomputer Centre of the
Russian Academy of Sciences.

\newpage

\begin{thebibliography}{}

\bibitem[Chirikov(1959)]{C59}
Chirikov, B.V. 1959, Atomnaya Energiya, 6, 630 [1960, J.\ Nucl.\
Energy Part~C: Plasma Phys., 1, 253]

\bibitem[Chirikov(1979)]{C79}
Chirikov, B.V. 1979, Phys. Rep., 52, 263

\bibitem[Deller \& Maddison(2005)]{DM05}
Deller, A.T., \& Maddison, S.T. 2005, Astrophys. J., 625, 398

\bibitem[Doyle et al.(2011)]{D11}
Doyle, L. et al., 2011, Science, 333, 1602

\bibitem[Duncan et al.(1989)]{DQT89}
Duncan, M., Quinn, T., \& Tremaine, S. 1989, Icarus, 82, 402

\bibitem[Gladman et al.(2012)]{G12}
Gladman, B. et al., 2012, Astron. J., 144, 23

\bibitem[Hairer et al.(1987)]{HNW87}
Hairer, E., N{\o}rsett, S.P., \& Wanner, G. 1987, Solving Ordinary
Differential Equations I. Nonstiff Problems (New York: Springer)

\bibitem[Holman \& Murray(1996)]{HM96}
Holman, M.J., \& Murray, N.W. 1996, Astron. J., 112, 1278

\bibitem[Holman \& Wiegert(1999)]{HW99}
Holman, M.J., \& Wiegert, P.A. 1999, Astron. J., 117, 621

\bibitem[Kouprianov \& Shevchenko(2003)]{KS03}
Kouprianov, V.V., \& Shevchenko, I.I. 2003, Astron. Astrophys., 410, 749

\bibitem[Kouprianov \& Shevchenko(2005)]{KS05}
Kouprianov, V.V., \& Shevchenko, I.I. 2005, Icarus, 176, 224

\bibitem[Lichtenberg \& Lieberman(1992)]{LL92}
Lichtenberg, A.J., \& Lieberman, M.A. 1992, Regular and Chaotic
Dynamics (New York: Springer)

\bibitem[Melnikov \& Shevchenko(1998)]{MS98}
Melnikov, A.V., \& Shevchenko, I.I. 1998, Sol. Sys. Res., 32, 480
(Astron.\ Vestnik, 32, 548)

\bibitem[Melnikov \& Shevchenko(2010)]{MS10}
Melnikov, A.V., \& Shevchenko, I.I. 2010, Icarus, 209, 786

\bibitem[Meschiari(2012)]{M12}
Meschiari, S. 2012, Astrophys. J., 752, 71

\bibitem[Morbidelli(2002)]{Morbi02}
Morbidelli, A. 2002, Modern Celestial Mechanics
(Taylor and Francis, Padstow)

\bibitem[Moriwaki \& Nakagawa(2004)]{MN04}
Moriwaki, K., \& Nakagawa, Y. 2004, Astrophys. J., 609, 1065

\bibitem[Mudryk \& Wu(2006)]{MW06}
Mudryk, L.R., \& Wu, Y. 2006, Astrophys. J., 639, 423

\bibitem[Murray \& Dermott(1999)]{MD99}
Murray, C.D., \& Dermott, S.F. 1999, Solar System Dynamics
(Cambridge: Cambridge Univ. Press)

\bibitem[Murray \& Holman(1997)]{MH97}
Murray, N.W., \& Holman, M.J. 1997, Astron. J., 114, 1246

\bibitem[Orosz et al.(2012a)]{O12a}
Orosz, J.A. et al., 2012, Astrophys. J., 758, 87

\bibitem[Orosz et al.(2012b)]{O12b}
Orosz, J.A. et al., 2012, Science, 337, 1511

\bibitem[Paardekooper et al.(2012)]{P12}
Paardekooper, S.-J., Leinhardt, Z.M., Th\'ebault, T., Baruteau, C.
2012, Astrophys. J., 754, L16

\bibitem[Pierens \& Nelson(2007)]{PN07}
Pierens, A., \& Nelson, R.P. 2007, Astron. Astrophys., 472, 993

\bibitem[Pierens \& Nelson(2008)]{PN08}
Pierens, A., \& Nelson, R.P. 2008, Astron. Astrophys., 483, 633

\bibitem[Quillen(2006)]{Q06}
Quillen, A.C. 2006, MNRAS, 365, 1367

\bibitem[Popova \& Shevchenko(2012)]{PS12}
Popova, E.A., \& Shevchenko, I.I. 2012, Astron.\ Lett., 38, 581
(Pis'ma Astron.\ Zhurnal, 38, 652)

\bibitem[Shevchenko(2002)]{S02c}
Shevchenko, I.I. 2002, in Asteroids, Comets, Meteors 2002,
ed. by Warmbein, B. (Berlin: ESA) 367

\bibitem[Shevchenko \& Kouprianov(2002)]{SK02}
Shevchenko, I.I., \& Kouprianov, V.V. 2002, Astron. Astrophys., 394, 663

\bibitem[Shevchenko \& Melnikov(2003)]{SM03}
Shevchenko, I.I., \& Melnikov, A.V. 2003,
JETP Lett., 77, 642 (Pis'ma ZhETF, 77, 772)

\bibitem[von Bremen et al.(1997)]{BUP97}
von Bremen, H.F., Udwadia, F.E., \& Proskurowski, W. 1997,
Physica D, 101, 1

\bibitem[Welsh et al.(2012)]{W12}
Welsh, W.F. et al., 2012, Nature, 481, 475

\bibitem[Wisdom(1980)]{W80}
Wisdom, J. 1980, Astron. J., 85, 1122

\bibitem[Wisdom(1987)]{W87}
Wisdom, J. 1987, Astron. J., 94, 1350

\end{thebibliography}
\end{document}